\documentclass[aps,prl,twocolumn,groupaddress,usenatbib,nofootinbib,showkeys,showpacs,
altaffilletter]{revtex4-1}

\begin{document}


\title{The rise of a tensor instability in Eddington-inspired gravity}

\author{Celia Escamilla-Rivera}
\affiliation{Fisika Teorikoaren eta Zientziaren Historia Saila, Zientzia 
eta Teknologia Fakultatea, Euskal Herriko Unibertsitatea, 644 Posta 
Kutxatila, 48080, Bilbao, Spain.}

\author{M\'aximo Ba\~nados}
 \affiliation{P. Universidad Cat\'olica de Chile, Avenida Vicuna Mackema 4860,
 Santiago, Chile}

\author{Pedro G. Ferreira}
\affiliation{Astrophysics, University of Oxford, DWB, Keble Road, Oxford, OX1 3RH, UK}


\begin{abstract}

In this work an extension to Eddington's gravitational action is analyzed. We consider the tensor perturbations
of a FLRW space-time in the Eddington regime in where the tensor mode is linearly unstable deep and
the resulting modifications to Einstein regime are quite strong.

\end{abstract}


\maketitle


\textit{Introduction}.-	
One of the greatest jigsaws in the current physics research is to understand the  starting point to general 
relativity. The Einstein-Hilbert action has been usually the initial core of gravitational theories, but an 
alternative idea was proposed by Eddington in Ref.~\cite{Eddington1923}, where the affine connection 
$\Gamma^{\mu}_{\phantom{\mu}\alpha\beta}$ is the fundamental field. Stylishly, the two suggestions
are equivalent by a factor of $\Lambda$. Moreover, Eddington proposal is incomplete because it doesn't 
include a matter sector. At this point an alternative theory of gravity was proposed in 
Ref.~\cite{Banados:2010ix} where the matter fields are introduced inside the gravitational action in a 
Palatini form
\begin{eqnarray}\label{palatini}
S_{EBI}[g,\Gamma,\Psi]&=& \frac{2}{\kappa}\int{d^4 x \left[\sqrt{\left|g_{\mu\nu}
+\kappa R_{\mu\nu}(\Gamma)\right|} -\lambda\sqrt{g}\right]}  \nonumber \\ &&
+ S_{m}[g,\Psi], 
\end{eqnarray}
where $\kappa =8\pi G$, $\Psi$ denotes any additional matter fields. $R_{\mu\nu}$ is the symmetric Ricci
tensor constructed with $\Gamma$. This action can reproduce  Eddington's original action at large values 
of $\kappa R$ and Einstein's at small values. Additionally, in this new theory the Eddington regime arise 
in the very early Universe and it may have led to a minimum scale factor. Hence, and more significantly, in this 
regime seems we can prevent the formation of cosmological singularities.
As an enthusiastic proposal there are several literatures that proof an interesting physics, for example in 
Ref.~\cite{Casanellas2012} was showed the possibility to test the Eddington corrections to Newtonian 
gravity using Solar physics and in Ref.~\cite{Pani:2011mg} the test was employed around compact 
rotating sources. The kindness of this new theory gives the possibility to reexpressed as a bigravity theory  
as was demonstrated in Ref.~\cite{Delsate2012}.


\textit{Instability in the Eddington Born-Infeld Theory}.-
\label{theory}
Our main point is to consider a perturbed homogeneous and isotropic space-time by choosing the following 
two metrics
\begin{eqnarray}\label{eq:metrics}
g_{\mu\nu}dx^\mu dx^\nu&=& -a^2 d\eta^2 +a^2\left(\delta_{ij} +h_{ij}\right) dx^i dx^j , \\
q_{\mu\nu}dx^\mu dx^\nu&=& -X^2 d\eta^2 +Y^2 \left(\delta_{ij}+\gamma_{ij}\right)dx^i dx^j ,
\end{eqnarray}
where $a$, $X$ and $Y$ are solely functions of conformal time, $\eta$. Here we shall work in the transverse
traceless (TT) gauge, which leaves only the tensor modes in perturbations, i.e ${\partial}_{i}h^{ij}={\partial}_{i}\gamma^{ij}=0$.
After straightforward calculations in where we found the background equations from Eq.(\ref{palatini})
and the perturbed field equations using Eq.(\ref{eq:metrics}) we have that even though the tensor perturbations 
in both the metric and the auxiliary metrics are multiplied by different conformal factors, they are identical in 
this theory as we presented in Ref.~\cite{EscamillaRivera:2012vz} . Furthermore, even in the Einstein 
regime, where $X=Y=a$, we find that $\gamma_{ij}$ is non trivial and completely locked to the behaviour of $h_{ij}$,
which evolution equation is given by 
\begin{eqnarray}
h_{ij}''+\left(3\frac{Y'}{Y}-\frac{X'}{X}\right)h_{ij}'+\left(\frac{X}{Y}\right)^2k^2h_{ij}=0. \label{gwave}
\end{eqnarray}
We now wish to see how this system evolves in the different regimes. In the Einstein regime we
find that the evolution is indistinguishable from Einstein gravity, even though the auxiliary metric
is perturbed and present. It is in the Eddington regime that we find novel behaviour  for a two different
values of $\kappa$:
\begin{itemize}
\item Case with $\kappa>0$. In Ref.~\cite{Scargill2012} it was shown a set of equations that
resurrects another of Eddington's ideas of doing away with  beginning and instead have the Universe indefinitely 
loitering in stasis in the distant past. To proceed we reexpress our background quantities in terms 
of conformal time, $\eta$ and the evolution equation for the tensor mode is
\begin{eqnarray}\label{kgwave}
h_{ij}''+2\alpha\frac{\exp(\alpha\Delta\eta)}{[1-\exp(\alpha\Delta\eta)]}h_{ij}'+\frac{\exp(\alpha\Delta\eta)}{[1-\exp(\alpha\Delta\eta)]}k^2h_{ij}=0. \quad\quad
\end{eqnarray}
Clearly we found an instability in the Eddington regime, in the asymptotic past when $\Delta\eta\rightarrow -\infty$.

\item Case with $\kappa <0$. We found that, if one choses a closed, positively 
curved spatial metric, it is possible to construct an oscillating (or \textit{Phoenix} Universe)\cite{Scargill2012} which undergoes an 
indefinite number of cycles. Such a model should, in principle allow us to study the evolution of perturbations 
through the various cycles and shed light on some of the issues that have been raised in the study of cyclic 
cosmologies. As we calculated in our previous work, Ref.~\cite{EscamillaRivera:2012vz}, the evolution
equation for the tensor mode is given by
 \begin{eqnarray}\label{kgwave}
 h_{ij}''+\frac{2}{\eta}h_{ij}'+\frac{k^2}{3\beta^2\eta^2}h_{ij}=0,  
 \end{eqnarray}
As in the case of previous case we find an instability this time at the bounce and the solutions blow up rendering 
such a space-time unstable to tensor mode perturbations.

\end{itemize}

\textit{Conclusions}.-
In this work we learn that traceless/transverse modes may play an unexpected role and should 
be included if possible in gravitational theories. Indeed by allowing more general perturbations it should 
also be possible trace the effect of nonlinear evolution of the tensor modes (coupled to radial modes) to search if 
the singularity can be stabilized in the nonlinear regime. It is remarkable to mentioned that exist a couple of 
analysis of perturbations in Ref.~\cite{Banados:2008fj} where did not find such any instability in the scalar 
sector and this might be a hint that 
it is the particular form of the Eddington inspired Born-Infeld theory that gives rise to such interesting behaviour. 

Last but not least, our analysis clearly hints at the possibility that interesting effects might arise in these 
theories in regions of density and curvature and gives us a new way of looking at the generation and
evolution of cosmological perturbations.


\textit{Acknowledgments}.-
C. E-R is grateful to M\'aximo Ba\~nados and Pedro G. Ferreira for proposing this problem and their devoted 
guidance and hospitality during the author's stay at the University of Oxford. Also thanks to R. Jantzen, 
K. Rosquist and R. Ruffini for their invitation and support in the MG13 meeting and Shinji Mukohyama chairman
in the Session AT4: Modify Gravity. The author is supported by Pablo Garc\'ia Fundation, FUNDEC, Mexico and
the Department of Theoretical Physics UPV/EHU Research Group 317207ELBE.




\end{document}